\documentstyle[epsfig,12pt]{article}
\newcommand{\la}[1]{\label{#1}}
\newcommand{\ur}[1]{(\ref{#1})}
\newcommand{\urs}[2]{(\ref{#1},\ref{#2})}

\newcommand{\eq}[1]{eq.~(\ref{#1})}
\newcommand{\eqs}[2]{eqs.~(\ref{#1},\ref{#2})}
\newcommand{\Eq}[1]{Eq.~(\ref{#1})}
\newcommand{\Eqs}[2]{Eqs.(\ref{#1}, \ref{#2})}

\newcommand{\half}{\frac{1}{2}}

\def\Tr{\mbox{Tr}}

\def\beq{\begin{equation}}
\def\eeq{\end{equation}}
\def\bea{\begin{eqnarray}}
\def\eea{\end{eqnarray}}

\begin{document}
\vskip 1.5true cm
\begin{center}
{\Large\bf Spontaneous breaking of color \\
\vskip .5true cm

in  ${\cal N}=1$ Super Yang--Mills theory without matter} \\

\vskip 1.5true cm

{\large\bf Dmitri Diakonov$^{\diamond *}$ and Victor Petrov$^*$} \\
\vskip .8true cm
$^\diamond$ {\it NORDITA, Blegdamsvej 17, DK-2100 Copenhagen \O,
Denmark} \\
\vskip .5true cm
$^*$ {\it St.Petersburg Nuclear Physics Institute, Gatchina 188 300,
Russia} \\
\vskip .2true cm
\end{center}
\vskip 1true cm
\begin{abstract}
\noindent
We argue that in the pure ${\cal N}=1$ Super Yang--Mills theory
gauge symmetry is spontaneously broken to the maximal Abelian
subgroup. In particular, colored gluino condensate is nonzero.
It invalidates, in a subtle way, the so-called strong-coupling
instanton calculation of the (normal) gluino condensate and
resolves the long-standing paradox why its value does not agree with
that obtained by other methods.
\end{abstract}
\vskip .2true cm
PACS: 11.15.-q, 11.15.Ex, 11.30 Pb
\vskip 1.5true cm 

\section{Introduction}

Pure ${\cal N}=1$ Super Yang--Mills (SYM) theory is known to
possess nonzero gluino condensate $<\!\lambda\lambda\!>$ whose phase
distinguishes between one of the discrete vacua of the theory
\cite{Wit1}. The exact value of $<\!\lambda\lambda\!>$ has been found
by several independent methods of controllable deformation to weak
coupling. One method \cite{ADS,NSVZ1} uses matter supermultiplets
whose nonzero Higgs condensate breaks explicitly the gauge group and
gives masses to certain fields. One is then able to compute
$<\!\lambda\lambda\!>$ from a single instanton, with extra fermion
zero modes contracted via mass terms. The other method \cite{DHKM}
compactifies the Euclidean space $R^4\to R^3\times S^1$, so that BPS
dyons arise as classical saddle points. Again, the gauge $SU(N)$
group is broken, this time spontaneously, by the nonzero expectation
value of the Yang--Mills potential $A_4$ in the compact direction.
Dyons have zero fermionic modes saturating $<\!\lambda\lambda\!>$;
it turns out to be independent of the circumference $l$ of the compact
dimension. It is then argued that the power of holomorphy allows one
to assert the same value in the decompactified limit $l\to\infty$.
We briefly review these methods below. The results of the two
seemingly different methods of getting $<\!\lambda\lambda\!>$
coincide, including the numerical coefficient \cite{DHKM}. The same
result which is apparently exact \cite{FP}, follows independently
from a deformation of the ${\cal N}=2$ theory -- see ref. \cite{RV}
for a recent discussion.\\

In all those approaches, the gauge symmetry is broken by the
deformation. Although in all cases the symmetry-breaking parameter
tends to zero as one approaches the strong coupling limit, one can
ask if the spontaneous breaking of gauge symmetry (i.e. a dynamical
Higgs effect) is not a property of the pure SYM theory itself. We
present arguments that it is indeed the case. Matter mutliplets which
break color explicitly or compactification which breaks it
spontaneously, serve as a `seed' to disclose the true nature of the
SYM theory in the strong-coupling limit. \\

Historically, the first calculation of the gluino condensate
\cite{NSVZ2} was directly in the strong-coupling limit of the
pure SYM theory. However, a seemingly ``clean'' calculation of
$<\!\lambda\lambda\!>^N$ in the pure $SU(N)$ SYM theory by
saturating it by instanton zero modes yields a value
different from the exact result. This paradox attracted much
attention over the years. There have been several attempts in the
past to explain the puzzle. It has been suggested \cite{AKMRV}
that instantons average over the $Z_N$ vacua, or over an additional
vacuum with zero gluino condensate \cite{KS}. However,
ref. \cite{RV} doubts the validity of those arguments. \\

In this paper, we suggest an alternative strong-coupling calculation
of the gluino condensate $<\!\lambda\lambda\!>^N$. On the one hand
it yields the correct result. On the other hand it is very close
to the old instanton calculation, and it becomes possible
to pinpoint what exactly is wrong there. Namely, the new calculation
reduces to the old strong-coupling instanton calculation
{\em provided one neglects long-ranged fields} vanishing as $1/l$ where
$l$ is the size of the system. Normally, such fields have no effect
on the local properties of the theory, but not in this case: a small
perturbation has a dramatic effect because the system is unstable
with respect to spontaneous color symmetry breaking. \\

\section{Spontaneous color symmetry breaking in the compactified SYM theory}

In this section we briefly review one of the ways to obtain the correct
value of the gluino condensate \cite{DHKM}. \\

Let us consider the $SU(2)$ SYM theory compactified to $R^3\times
S^1$ with the `time' dimension $x_4$ being of circumference $l$.
It should be stressed that it is {\em not} an introduction of the
physical temperature $T=1/l$ as fermions satisfy periodic conditions
in the $x_4$ direction. Therefore, the usual perturbative periodic
potential in $A_4$ does not emerge as in the temperature case: owing
to supersymmetry it is zero to all orders of the perturbation theory.
We remind the reader that the perturbative potential $V(A_4)$ is zero
at $\sqrt{A_4^aA_4^a}= 0,2\pi/l,4\pi/l\ldots $ at which points the
Polyakov line (the holonomy) is trivial. If the holonomy is
nontrivial (more precisely, if its spatial average is nontrivial) then
$V(A_4)> 0$ and the corresponding gauge configuration has an
unacceptable volume-divergent positive energy. This is the usual
argument against configurations with nontrivial average holonomy in
the pure gauge theory \cite{GPY}. However, in the compact SYM theory
the perturbative potential is identically zero for any $A_4$ and one
is free to consider configurations with any holonomy at spatial infinity.\\

Choosing the gauge where at spatial infinity $A_4\to v\tau^3/2$
one finds that there are two self-dual ($L,M$) and two
anti-self-dual ($\bar L, \bar M$) dyon solutions of the YM equations,
with the same asymptotic value of $A_4^3=v$ at spatial infinity
\cite{LY,DHKM}. These solutions have all four possible signs of the
electric and magnetic charges. The corresponding fields are given
explicitly in the Appendix. \\

The nonperturbative dyon-induced superpotential found in ref.
\cite{DHKM} shows that the minimum (zero) energy is achieved when
the weights of the $L$ and $M$ dyons become equal, which happens at

\beq
\sqrt{A_4^aA_4^a}=v=\frac{\pi}{l}.
\la{vev}\eeq
We notice that this value corresponds to the maximum of the would-be
perturbative potential but it is absent. The system settles at
the minimum \ur{vev} of the nonperturbative potential.
It clearly demonstrates that in compactified SYM theory color is
spontaneously broken by the Higgs mechanism, with $A_4^a$ playing
the role of the Higgs field in the adjoint representation. The
symmetry breaking pattern is $SU(2)\to U(1)$. For higher $SU(N)$
gauge groups the minimum (zero) energy is achieved at \cite{DHKM}

\beq
A_4={\rm diag}\left(\frac{N-1}{N},\frac{N-3}{N},\ldots,-\frac{N-1}{N}
\right)\frac{\pi}{l}.
\la{vevN}\eeq
It means that the $SU(N)$ gauge group is spontaneously
broken down to the maximal Abelian subgroup $U(1)^{N-1}$, at least
at small compactification circumference $l\ll \Lambda$ where
$\Lambda$ is the SYM scale parameter. \\

\Eq{vevN} is not gauge-invariant. To put it in a gauge-invariant form 
one can consider the Polyakov line (the holonomy) along the compactified 
dimension; its eigenvalues are gauge invariant:
\bea
\la{P} 
\!\!P\!\!&\!\!=\!\!&\!{\mbox P}\,
\exp\left(i\!\int_0^l\!\!dx^4A_4\right)\!
=\!{\rm diag}\left(\exp\left(i\pi\frac{N\!-\!1}{N}\!\right),
\exp\left(i\pi\frac{N\!-\!3}{N}\!\right)...
\exp\left(\!-i\pi\frac{N\!-\!1}{N}\!\right)\!\right)\!,
\\
\la{TrP}
\Tr\,P\!&\!=\!&\!0.
\eea
For $SU(2)$ the Polyakov line's eigenvalues are
\beq
P={\rm diag}\left(i,-i\right),\qquad \Tr P=0.
\la{P2}\eeq

One dyon can be considered in whatever gauge. However, if we wish 
to consider the vacuum filled by dyons, we have to take more than 
one dyon. Two and more dyons can be put together only in the singular 
`stringy' gauge (see Appendix) where all of them have the same orientation 
in color space. This orientation is preserved throughout the $R^3$ volume. 
The mere notion of the ensemble of dyons (or monopoles) implies that color 
symmetry is broken. Of course, once color is aligned, one can always 
randomize the color orientation by an arbitrary point-dependent gauge 
transformation, just as the direction of the Higgs field can be randomized
but that does not undermine the essence of the Higgs effect. In our case,
the eigenvalues of the holonomy \urs{P}{P2} and $\Tr P=0$ are gauge-invariant 
signatures of the Higgs effect. \\

\Eqs{TrP}{P2} do not mean that $\Tr P$ is zero identically: 
it experiences point-to-point fluctuations, of course. For example,
if $\Tr P$ is measured near the dyon center it will be anything but zero. 
The statement is that $\Tr P\to 0$ far away from dyon centers. 
A simple calculation shows that also $<\!\Tr P\!>=0$ for a Coulomb gas
of dyons. As a matter of fact, this is the usual confinement requirement. \\

Although $A_4=\frac{\pi}{l}\to 0$ in the strong-coupling decompactified 
limit $l\to\infty$, taken naively, the holonomy \ur{P} remains non-trivial. 
Unfortunately, it is not a holomorphic quantity so that one cannot prove
it rigorously. Nevertheless, we shall argue in the next section that the 
holonomy does remain non-trivial and that color symmetry remains 
broken in the decompactified limit. To that end we would need to consider 
the gluino condensate which is a holomorphic quantity. \\

Both $L$ and $M$ dyons have two gluino zero modes being the
Grassmann partners of the four translational zero modes and thus
being related to the dyon field strength:

\beq
\lambda^{a\alpha}_{\rm zero\;mode}
=\left(\sigma^+_\mu\right)^\alpha_{\dot\beta}
\left(\sigma^-_\nu\right)^{\dot\beta}_\gamma\xi^\gamma F_{\mu\nu}^a
=\left(\sigma_i\right)^\alpha_\gamma\xi^\gamma E_i^a
\la{gzm1}\eeq
where $E_i^a=B_i^a$ is the electric field strength of a dyon, see
\eqs{EiM}{EiL}. As shown in ref. \cite{DHKM} the dyon zero modes 
saturate the gluino condensate

\bea
\nonumber
<\!\lambda\lambda\!>&=&
<\epsilon_{\alpha\beta}\lambda^{a\alpha}(x)
\lambda^{a\beta}(x)>
=2\,\frac{\Lambda^3}{4\pi v}\int\!d^3z\,E^a_i(x-z)E^a_i(x-z) \\
\nonumber\\
\la{gc1}
&=&\frac{\Lambda^3}{v}\int_0^\infty\!drr^2\left[2F_1^2(r)+F_2^2(r)\right]
=\Lambda^3\equiv\frac{16\pi^2\,M_{\rm PV}^3}{g^2(M_{\rm PV})}
\,\exp\left[-\frac{4\pi^2}{g^2(M_{\rm PV})}\right]
\eea
where $\Lambda$ is the renormalization-invariant combination of the
Pauli-Villars regularization mass and the bare gauge coupling 
\footnote{$\Lambda^3$ used here is 6 times bigger than that
used in the QCD convention.}. 
The coefficient `2' comes from summing up the (equal) contributions
of $L$ and $M$ dyons. The radial functions $F_{1,2}(r)$ are the
profile functions of the dyon, see \eqs{F1}{F2}. 
We remark that it is actually the {\em anti}-self-dual $\bar L,\bar M$ dyons that
lead to the $<\!\lambda\lambda\!>$ condensate 
(self-dual $L,M$ dyons lead to  $<\!\bar\lambda\bar\lambda\!>$)
but we shall not stress this distinction. Although technically obtained 
in the small $l$ limit the result \ur{gc1} coincides with the exact one in the
decompactified strong-coupling limit. \\

\newpage
\section{Instantons vs dyons}

Let us now recall the strong-coupling instanton calculation of the
gluino condensate \cite{NSVZ2}. Contrary to the dyon, the instanton
has four gluino modes for the $SU(2)$ group. Therefore, a single
gluino condensate cannot be saturated by an instanton. Instead, one
considers a two-point correlation function

\beq
C(x-y)=<\epsilon_{\alpha\beta}\lambda^{a\alpha}\lambda^{a\beta}(x)
\,\,\,
\epsilon_{\gamma\delta}\lambda^{c\gamma}\lambda^{c\delta}(y)>
\la{C1}\eeq
which can be saturated by a single instanton. This correlation
function does not actually depend on $x\!-\!y$ owing to
supersymmetry. Therefore, one can evaluate the correlator at
$|x\!-\!y|\to 0$ using small-size instantons. Since the correlator is
$|x\!-\!y|$-independent, the same value holds at $|x\!-\!y|\to\infty$
where it can be factorized into the product of two gluino condensates
$<\!\lambda\lambda\!>$. This procedure known as `strong-coupling
instantons' gives a famous discrepancy factor of $\frac{4}{5}$ as
compared to the exact result. We shall show that the evaluation of
$C(x-y)$ from an instanton is incorrect both for vanishing and for
large $|x\!-\!y|$:  the seemingly clean calculation has a loophole
because of the spontaneous breaking of the gauge group. \\

We start with a simple algebraic argument showing that instantons
do not handle color in a way compatible with supersymmetry.
Let us consider the correlation function of two gauge-invariant
gluino bilinears like in \eq{C1} but which are not contracted in
spinor indices. Since fermion operators anticommute we find that the
correlation function must be antisymmetric inside the two pairs of
spin indices:

\beq
<\lambda^{a\alpha}\lambda^{a\beta}(x)\,\,\,
\lambda^{c\gamma}\lambda^{c\delta}(y)>=\frac{1}{4}\,\epsilon^{\alpha\beta}
\epsilon^{\gamma\delta}\,C(x-y), \qquad C={\rm const.}
\la{C2}\eeq
This correlator is actually $|x\!-\!y|$-independent since its
contraction with $\epsilon_{\alpha\beta}\epsilon_{\gamma\delta}$ is.
Therefore one can put $x=y$ in \eq{C2} so that it becomes a one-point
average. We next consider a one-point average of gluino fields which
are contracted in spin but not in color indices:

\beq
T^{ab,cd}=<\epsilon_{\alpha\beta}\lambda^{a\alpha}\lambda^{b\beta}(x)\,
\epsilon_{\gamma\delta}\lambda^{c\gamma}\lambda^{d\delta}(x)>.
\la{T1}\eeq
Under gauge transformations this tensor is gauge-rotated with respect 
to all indices. After averaging over gauge rotations only invariant 
tensors can result. Fermion statistics requires that $T^{ab,cd}$ is 
symmetric in $(ab)$ and in $(cd)$. In the $SU(2)$ gauge theory there 
are only two possible invariant structures made of Kronecker deltas, 
consistent with symmetry:

\beq
T^{ab,cd}=A\,\delta^{ab}\delta^{cd}+B\,(\delta^{ac}\delta^{bd}
+\delta^{ad}\delta^{bc}).
\la{T2}\eeq
In higher groups more structures are possible but we do not consider
them here. Contracting \eq{T1} once with $\delta^{ab}\delta^{cd}$
and the other time with $\delta^{ac}\delta^{bd}$ we reduce it to
\eq{C2} contracted in the first case with
$\epsilon_{\alpha\beta}\epsilon_{\gamma\delta}$ and in the second
case with $-\epsilon_{\alpha\gamma}\epsilon_{\beta\delta}$ (the minus
sign arises from adjusting the order of fermion operators). It gives
a system of linear equations on the coefficients $A,B$:

\beq
\left\{\begin{array}{ccccc}
9A & + & 6B  & = & C\\
3A & + & 12B & = & -\half C
\end{array}
\right.
\la{sys}\eeq
with a unique solution

\beq
A=\frac{C}{6},\qquad B=-\frac{C}{12}.
\la{AB}\eeq
Thus, the color structure of the one-point average \ur{T1} is
unambiguously determined by supersymmetry:

\beq
<\epsilon_{\alpha\beta}\lambda^{a\alpha}\lambda^{b\beta}(x)\;
\epsilon_{\gamma\delta}\lambda^{c\gamma}\lambda^{d\delta}(x)>
=\frac{C}{6}\left[\delta^{ab}\delta^{cd}-\frac{1}{2}(\delta^{ac}\delta^{bd}
+\delta^{ad}\delta^{bc})\right].
\la{T3}\eeq

The next observation is that the instanton contribution to the l.h.s. of
\eq{T3} fails to reproduce its color structure. There are four gluino
zero modes in the instanton background: two are super-translational
and two are super-dilatational (or super-conformal) \cite{NSVZ2}.
One has to insert those zero modes into \eq{T3}, in all possible
combinations. A simple exercise in algebra demonstrates that
only the color-singlet structure $\delta^{ab}\delta^{cd}$ arises,
with the coefficient $B$ being identically zero! This is true
not only for exactly coinciding points $x=y$ but also for $x\neq y$.
It is true identically, even before one integrates over instanton
center and sizes. We have also checked that it does not depend
on the gauge in which the instanton field is considered. \\

To gain further insight, let us introduce a traceless color gluino bilinear 
operator

\beq
\Lambda^{ab}=\epsilon_{\alpha\beta}
\left(\lambda^{a\alpha}\lambda^{b\beta}
-\frac{\delta^{ab}}{N^2-1}\,\lambda^{e\alpha}\lambda^{e\beta}\right),
\qquad\quad \Lambda^{aa}= 0.
\la{glop}\eeq
In the case of the $SU(2)$ group $\Lambda^{ab}$ belongs to the
irreducible dimension-5 `isospin'-2 representation. For higher $N$
the symmetric rank-2 traceless representation is reducible; for
example in $SU(3)$ it is a mixture of ${\bf 8}^d$ (adjoint) and
${\bf 27}$ representations. \\

A direct consequence of \eq{T3} is that the one-point average 
of gluinos in the traceless dimension-5 representation of $SU(2)$ is

\beq
<\Lambda^{ab}(x)\,\Lambda^{ab}(y)>_{|x-y|\to 0}\quad
\longrightarrow \quad -\frac{5}{6}\,<\!\lambda\lambda\!>^2.
\la{LaLa}\eeq
In the $SU(N)$ case $-\frac{5}{6}$ is replaced by the general
$-\half\frac{N^2+1}{N^2-1}$; the negative sign is related to that
$\Lambda^{ab}$ is a fermion operator. Meanwhile, the strong-coupling
instanton calculation of this quantity (implying $B=0$) yields
identical zero! Instantons are `color-blind' and average out any
`colored' operator. Were color symmetry preserved in the
pure SYM theory strong-coupling instantons would be all right.
This is the first indication that the instantons' failure is related to the 
actual color symmetry breaking in the theory, but there will be more 
\footnote{It was noticed earlier \cite{HKLM} that multi-instantons do 
not support the cluster decomposition of gluino correlators. 
Since \eq{LaLa} is very general (it can be derived directly from 
first rearranging gluinos into color-singlet operators and then applying 
\eq{C2}), instantons' failure to reproduce the equation is another but simpler
manifestation of the non-clusterization.}. \\

We next consider the correlation function \ur{C1} at large
separations between $x$ and $y$. This correlation function has
chirality two, meaning that only gauge configurations with unity
topological charge can contribute. Instanton is an obvious candidate.
From the instanton viewpoint, the correlator is saturated by
instantons of size $\rho\sim |x-y|$ \cite{NSVZ2} but the result turns
out to be $\frac{4}{5}$ of the exact one. So far all calculations yielding 
the correct value were made for a single gluino condensate,  
whereas the suspicious strong-coupling instanton calculation was 
for the two-point correlator. Therefore, to pin down the mistake one should 
perform in parallel a correct calculation but for the two-point correlator
of gluino condensates. \\

There is an alternative strong-coupling calculation of
$<\!\lambda\lambda(x)\, \lambda\lambda(y)\!>$ stemming from the
compactified version of the SYM theory. The unity topological charge
can be obtained from any two dyons $LL,MM,LM$. In the compactified
$R^3\times S^1$ space there are exact classical solutions of all
three types. The full eight-parameter static double-monopole $MM$
solution has been known for a while \cite{FHP}. The $LL$
double-monopole solution can be obtained from the $MM$ one by a gauge
transformation. The time-dependent eight-parameter $LM$ solution has
been recently constructed explicitly and named `the caloron with
non-trivial holonomy' \cite{KvB,LL}. The first two objects have
double electric and magnetic charges so that both their electric and
magnetic fields decay as $1/r^2$ at large distances. The third
object has zero charges so that it is similar to the instanton. \\

To compute the correlator $<\!\lambda\lambda(x)\, \lambda\lambda(y)\!>$
one needs to take one of the three ($LL$, $MM$, $LM$) exact
solutions, find their four adjoint fermion zero modes, substitute
them into the correlator in question in all possible combinations,
and integrate over the solutions' moduli space; finally sum up the
contributions of all three exact solutions. \\

In practice, the calculation of the correlator depends on the relation
between $|x\!-\!y|$ and the compactification circumference $l$. Let
us first discuss the `weak-coupling' case of $l\ll |x\!-\!y|$. 
In this case, only part of the moduli space of the exact solutions 
contribute, corresponding to widely separated `constituent' $L,M$ 
dyons.  Since the field of constituents decreases rapidly beyond 
their size $\sim l$, the leading contribution comes from one of the
dyons staying at the distance $\sim l$ from point $x$ and the other
being at the distance $\sim l$ from point $y$. Their interference
can be neglected. Therefore, at $l\ll |x\!-\!y|$ the calculation of
the correlator just copies (twice) the calculation of the gluino
condensate from a single dyon \cite{DHKM} (recall section 2) with an
evident result: the correlator is independent of $|x\!-\!y|$ and
coincides with the square of the (correct) gluino condensate. Notice
that all four possible combinations $LL,MM,LM,ML$ contribute exactly
$\frac{1}{4}<\!\lambda\lambda\!>^2$ apiece. \\

We next turn to the opposite case, $l\gg |x\!-\!y|\gg 1/\Lambda$,
appropriate to the decompactified strong-coupling limit. One
should keep in mind that compactification does not spoil
supersymmetry. Owing to supersymmetry $<\!\lambda\lambda(x)\,
\lambda\lambda(y)\!>$ is independent of $|x\!-\!y|$ for any given
$l$. Thus, the correlator must be precisely the same as in the
previous case and equal to the square of the gluino condensate. The
correct result can be foreseen without calculations!  \footnote{Out
of curiosity, we have computed the one-point average \ur{T1} assuming
a sum Ansatz of dyons at all separations. Surprisingly, it works quite
well: the color structure \ur{AB} following from supersymmetry is
reproduced and the absolute value of the gluino condensate turns out
to be only 4\% bigger than the exact one. It would be illuminating to
compute $<\!\lambda\lambda(x)\, \lambda\lambda(y)\!>$ exactly from
the $LL(MM)$ and $LM$ solutions.} \\

When $l\to\infty$ while $|x\!-\!y|$ is kept fixed, the exact
$LL(MM)$ and $LM$ solutions look very different. The $LM$ solution
(the caloron) at $l\to\infty$ and fixed size $\rho$ becomes the
instanton \cite{KvB,LL}. Its action density is well localized both
in $x_4$ and space. In the leading order in $1/l$ the solution is
the usual instanton. The difference with the instanton shows up
only in the subleading $1/l$ terms. As to the $LL$ and $MM$ exact
solutions, they can be made static by an appropriate gauge choice.
At $l\to \infty$ their field is weak everywhere: it is of the order of
$1/l$ inside the region of space $\sim l$ and falls as $1/r$ outside
that region. The action gets its unity value owing to the integration
of a weak field over a large volume.\\

Naively, one would argue that fields of the order of $1/l\to 0$
are irrelevant for the calculation of the gluino condensate which
is a local quantity, and hence one would {\it i}) neglect altogether
the $LL$ and $MM$ contributions, {\it ii}) replace the exact $LM$
field by the instanton. Following this argument, one would conclude
that the two-dyon and the instanton calculations are equivalent
in the strong-coupling limit. However, we shall see in a moment that
this is incorrect. \\

The $LL,LM,MM$ solutions represent sectors with definite
(electric, magnetic) charges (-2,-2), (0,0) and (2,2), respectively.
These sectors do not mix up under supersymmetric transformations:
it is only their moduli spaces that transform (separately) under
supersymmetry. It means that the independence of the correlator
$<\!\lambda\lambda(x)\, \lambda\lambda(y)\!>$ of $|x\!-\!y|$ is
satisfied separately for the three sectors. At $l\ll |x\!-\!y|$ we
know that $LL$ and $MM$ sectors contribute apiece exactly
$\frac{1}{4}$ of the gluino condensate squared, whereas the $LM$
sector contributes exactly the other half. Because of supersymmetry,
at $l\gg |x\!-\!y|$ those configurations contribute precisely the
same fractions, despite that the $LL,MM$ fields tend to zero. At the
same time the exact $LM$ configuration (i.e. the caloron with
non-trivial holonomy) contributes precisely $\half$ of the gluino
condensate squared, whereas the instanton (to which it is reduced if
one neglects $1/l$ corrections) is known to contribute $\frac{4}{5}$.
Rather unusual, a vanishing field $A_4\sim 1/l$ is necessary to
maintain the correct result for the local gluino condensate
\footnote{Experts in the Schwinger model may find the present
situation familiar: the chiral condensate in the $2d$ QED is induced
by fields that are of the order of $1/l$ where $l$ is the size of the
world \cite{DDP}.}. \\

The difference between strong-coupling instanton and multi-dyon
calculations becomes even greater for higher $SU(N)$ groups.
At large $N$, the instanton gives only a $O(1/N)$ fraction of the true 
gluino condensate (see below); where does the rest come from? \\ 

In the compactified $SU(N)$ gauge theory, there are $N-1$ `static'
dyons $M_1...M_{N-1}$ having unit (electric, magnetic) charges
with respect to the $N\!-\!1$ Abelian subgroups, and one
`time-dependent' $L$ dyon \cite{LY,KvB}. When one computes
the single gluino condensate in compactified space each of the $N$
configurations contributes equally $\frac{1}{N}<\!\lambda\lambda\!>$
\cite{DHKM}. Adding up the contributions of $M_1,M_2...M_{N-1}$ and 
$L$ dyons one gets the correct gluino condensate. \\

To compare it with the strong-coupling instanton calculation, one
considers a $N$-point correlator

\beq
<\lambda\lambda(x_1)\,\lambda\lambda(x_2)\ldots
\lambda\lambda(x_N)> \quad\longrightarrow\quad
<\!\lambda\lambda\!>^N,
\la{lambdaN}\eeq
with $x_1\ldots x_N$ taken far apart. This correlator can be saturated by
one instanton but also, in the compactified space, by exact N-dyon
solutions. Again, we start with the case $l\ll |x_{mn}|$ where the exact 
multi-dyon solutions reduce to widely separated constituents. Each of the 
$N$ dyon species stay at the distance $\sim l$ from the points $x_1\ldots x_N$, 
with all possible $N^N$ permutations. Therefore, the l.h.s. of \eq{lambdaN} is

\beq
\left(\frac{1}{N}<\!\lambda\lambda\!>\right)^N\,N^N
=<\!\lambda\lambda\!>^N
\la{corN}\eeq
as it should be. By supersymmetry, the l.h.s of
\eq{lambdaN} does not depend on the relation between the
compactification circumference $l$ and the separations $x_{mn}$.
Therefore, the same result holds at $l\gg |x_{mn}|$, i.e. in the
strong-coupling limit. Meanwhile, only one particular configuration,
namely $M_1M_2...M_{N-1}L$, has zero (electric, magnetic) charges
with respect to all $U(1)$ subgroups. It is the `caloron with
non-trivial holonomy' of refs. \cite{LY,KvB}. At $l\to \infty$ it
becomes the usual instanton of the $SU(N)$ gauge group, {\em plus}
$1/l$ corrections. The instanton contribution to the
l.h.s. of \eq{lambdaN} is \cite{NSVZ2,AKMRV,FS,FP}

\beq
\left(<\!\lambda\lambda\!>_{\rm inst}\right)^N\quad
=\quad\frac{2^N}{(N-1)!\,(3N-1)}\,
<\!\lambda\lambda\!>^N\quad\stackrel{N\to\infty}{\longrightarrow}
\quad\left(\frac{2e}{N}\,<\!\lambda\lambda\!>\right)^N.
\la{corNinst}\eeq
Meanwhile, the true caloron contribution to \eq{lambdaN} is

\beq
\left(<\!\lambda\lambda\!>_{\rm calor}\right)^N\quad
=\quad\left(\frac{1}{N}<\!\lambda\lambda\!>\right)^N\,N!
\quad\stackrel{N\to\infty}{\longrightarrow}
\quad\left(\frac{1}{e}\,<\!\lambda\lambda\!>\right)^N
\la{corNcalor}\eeq
where the factor $N!$ comes from the permutations of $M_1,M_2,...,L$.
Again, we see that a vanishing distinction between the instanton
and the caloron with non-trivial holonomy makes a big difference.
Also, contributions from all the rest of multi-dyon solutions
whose field at $l\to \infty$ is $O(1/l)$ (or less) everywhere,
are needed to maintain the correct result for the gluino
condensate in the strong-coupling limit. \\

We see, thus, that the non-trivial holonomy is important
in determining the gluino condensate in the strong-coupling regime.
This is not very usual. The holonomy is a global quantity;
the difference between trivial and non-trivial holonomy is the
difference between $A_4=0$ and $A_4=\pi/l\to 0$. The fact that
this tiny difference plays a crucial role in determining such a local
quantity as $<\!\lambda\lambda\!>$ means that the system is unstable
with respect to infinitesimal perturbation breaking color symmetry.
In other words, the gauge group is spontaneously broken.

\section{Color gluino condensate from a deformation of ${\cal N}=2$
theory}

In this section we compute directly the value of the color-breaking gluino 
condensate $\Lambda^{ab}$, see \eq{glop}. 
To that end, we consider the compactified version of the
${\cal N}=2$ theory. As compared to the pure SYM theory, it has an
additional chiral multiplet $(\Psi^{a\alpha},\Phi^a)$ in the adjoint
representation. \\

The classical potential $g^2\Tr[\bar \Phi\Phi]^2$ has a flat
zero-energy valley which we shall choose in the form

\beq
\Phi^a=\left(\begin{array}{c}0\\0\\V\end{array}\right)
\la{val}\eeq
where $V$ is an arbitrary complex number. It breaks the color group
$SU(2)\to U(1)$ even without compactification. However, we shall add
the mass term for the chiral supermultiplet,

\beq
m\left(\epsilon_{\alpha\beta}\Psi^{a\alpha}\Psi^{a\beta}
+\Phi^a\Phi^a\right).
\la{mass}\eeq
In the decompactified case the mass term drives $V\to 0$ at
large $m$ \cite{SW,GVY}. At large $m$ the matter supermultiplet
decouples and one is left with the pure SYM theory with a seemingly
restored full $SU(2)$ gauge group. Such conclusion is, however, too
hasty. Compactification of the softly broken (by the mass term)
${\cal N}=2$ theory is a way to make a gradual transition to the SYM
theory, ultimately in the strong coupling regime. We shall see that
the $SU(2)$ group is not restored in that limit but remains broken to
$U(1)$ by the colored gluino condensate.\\

Compactifying the $x_4$ coordinate one finds classical solutions
being $L,M$ dyons modified by the presence of the scalar field
$\Phi^a$. Assuming the fields are `time'-independent and $\Phi^a$ is
parallel to $A_4^a$, the modified $M$ dyon in the regular `hedgehog'
gauge is given by (cf. \eqs{A41}{Ai1})
\bea
\la{Phi}
\Phi^a &=& -n_a\,V\Phi\left(\sqrt{v^2+V^2}\,r\right),\qquad
\Phi(z)=\coth z -\frac{1}{z} \\
\nonumber\\
\la{A4p}
A_4^a &=& -n_a\,v\,\Phi\left(\sqrt{v^2+V^2}\,r\right)
\quad \stackrel{z\to\infty}{\longrightarrow}\quad
-n^a\left(v-\frac{v}{\sqrt{v^2+V^2}}\frac{1}{r}\right),\\
\nonumber \\
\la{Aip} A_i^a &=& \epsilon_{aij}n_j\,
\frac{1-R\left(\sqrt{v^2+V^2}\,r\right)}{r},
\qquad R(z)=\frac{z}{\sinh z}.
\eea
Its action is $\frac{4\pi l}{g^2}\sqrt{v^2+V^2}$. The $S_-$ gauge
transformation \ur{Sm} puts the $\Phi^a,A_4^a$ fields along
the third color axis, with the asymptotics
\bea
\la{asPhiM}
\Phi^a
&\simeq & \delta^{a3}\left(V-\frac{V}{\sqrt{v^2+V^2}}\,
\frac{1}{r}\right),\\
\la{asA4M}
A_4^a
&\simeq & \delta^{a3}\left(v-\frac{v}{\sqrt{v^2+V^2}}\,
\frac{1}{r}\right).
\eea

The $L$ dyon is obtained by replacing $v\to \frac{2\pi}{l}-v$ and
$V\to -V$ in eqs.(\ref{Phi}-\ref{Aip}). It is then transformed to
the stringy gauge by $S_+$ \ur{Sp} and subsequently gauge-transformed
by the time-dependent matrix $U(x^4)$ \ur{U}. The fields' asymptotics
become
\bea
\la{asPhiL}
\Phi^a &\simeq &
\delta^{a3}\left(V-\frac{V}{\sqrt{\left(\frac{2\pi}{l}-v\right)^2+V^2}}
\,\frac{1}{r}\right),\\
\la{asA4L}
A_4^a &\simeq &
\delta^{a3}\left(v+\frac{\frac{2\pi}{l}-v}
{\sqrt{\left(\frac{2\pi}{l}-v\right)^2+V^2}}
\,\frac{1}{r}\right).
\eea
The action of the modified $L$ dyon is
$\frac{4\pi l}{g^2}\sqrt{\left(\frac{2\pi}{l}-v\right)^2+V^2}$.\\

Both $L$ and $M$ dyons have two $\lambda$ and two $\Psi$ zero modes.
The mass term for $\Psi$ allows one to contract the $\Psi$ zero modes
of a dyon.  The $L,M$-induced superpotential is a slight
modification of that found in ref. \cite{DHKM} in the pure SYM case:

\beq
{\cal W}_{\rm dyon}=\left(M_{\rm PV}^{{\cal N}=2}\right)^2\,m\,
\left[\exp\left(-\frac{4\pi l}{g^2}\sqrt{v^2+V^2}\right)
+\exp\left(-\frac{4\pi l}{g^2}\sqrt{\left(\frac{2\pi}{l}-v\right)^2
+V^2}\right)\right]
\la{sp}\eeq
Here the Pauli--Villars mass of the full ${\cal N}=2$ theory
appears in the second power since there are four boson and four
fermion zero modes. The factor $m$ arises from the contraction
of $\Psi$ zero modes via the mass term. Apparently, the minimum
(zero) energy is achieved, as before, at $v=\frac{\pi}{l}$,
independently of the v.e.v. of the matter field $\Phi^a$. From now
on, we shall use this value of $<\!A_4\!>$. 
\\

It is straightforward to calculate the gluino condensate in this
setting, basically repeating the steps leading to \eq{gc1}. The only
(technical) difference is that the dyon weight is now proportional
to $\left(M_{\rm PV}^{{\cal N}=2}\right)^2\,m$. This quantity is,
however, equal to $\left(M_{\rm PV}^{{\cal N}=1}\right)^3$ at large
$m$ (see e.g. \cite{RV}). Therefore, we get the same result as
before:

\beq
<\!\lambda\lambda\!>=<\!\lambda^1\lambda^1\!>+<\!\lambda^2\lambda^2\!>
+<\!\lambda^3\lambda^3\!>=\Lambda^3.
\la{gc2}\eeq

\vskip .5true cm
The ${\cal N}=2$ extension allows us to compute another holomorphic
quantity

\beq
\chi=
\left\langle\frac{\Lambda^{ab}\Phi^a\Phi^b}{\Phi^c\Phi^c}\right\rangle
\la{LPP}\eeq
where $\Lambda^{ab}$ is the traceless gluino bilinear in the dimension-5
representation, see \eq{glop}. This operator is chiral and transforms
under supersymmetry through the parameter $\epsilon$ only (not
$\bar\epsilon$). It gets a contribution from one dyon but
cannot acquire corrections either from perturbation theory or from
additional dyon pairs. Therefore, we can find the above average
at small $l$ and claim that it remains unaltered in the
decompactified limit, just as the normal gluino condensate
\ur{gc2} does.\\

Saturating $\Lambda^{ab}$ by the two gluino modes of a dyon we obtain
(cf. \eq{gc1}):

\beq
\Lambda^{ab}\rightarrow E^a_iE^b_i
-\frac{\delta^{ab}}{3}E^e_iE^e_i
=\left(F_2^2(r)-F_1^2(r)\right)
\left(\delta^{a3}\delta^{b3}-\frac{1}{3}\delta^{ab}\right).
\la{La}\eeq
\Eq{La} is written for $M$-dyons; in the case of the time-dependent
$L$-dyons \eq{La} should be gauge-rotated by a time-dependent matrix
\ur{U}. It is easy to check, however, that this gauge transformation commutes
with the color structure in \eq{La}, i.e. leaves it unchanged.
Therefore, \eq{La} is correct both for $L$ and $M$ dyons. \\

The scalar field $\Phi^a$ is directed along the third
color axis, according to \eqs{asPhiM}{asPhiL}. The concrete profile
of the $\Phi^a$ solution cancels out in the ratio
\ur{LPP}. For the same reason the v.e.v. of the $\Phi$ field is
irrelevant also, although it goes to zero at large $m$ (necessary
to pass from the ${\cal N}=2$ to the pure SYM theory). It is only the
color direction of $\Phi^a$ that matters in \eq{LPP}, but it
is fixed by the dyon solution. Consequently, the average \ur{LPP} is
\bea
\nonumber
\chi 
&=&<\!\Lambda^{33}\!>
=\frac{2}{3}<\!\lambda^3\lambda^3\!> -\frac{1}{3}<\!\lambda^1\lambda^1\!>
-\frac{1}{3}<\!\lambda^2\lambda^2\!>\\
\la{chi}
&=&\frac{2}{3}\,\frac{\Lambda^3}{4\pi v}\int\!
d^3z\left[F_2^2(r)-F_1^2(r)\right]
=\frac{2}{3}\,\Lambda^3\left(2-\frac{\pi^2}{6}\right).
\eea
Combining it with \eq{gc2} we find
\bea
\la{12}
<\lambda^1\lambda^1>&=&<\lambda^2\lambda^2>\quad=\quad
\Lambda^3\int\!dr\,r^2F_1^2(r)=\Lambda^3\frac{\pi^2-6}{18}
\simeq 0.214978\,\Lambda^3,\\
\la{3}
<\lambda^3\lambda^3>&=&
\Lambda^3\int\!dr\,r^2F_2^2(r)\quad=\quad\Lambda^3\frac{15-\pi^2}{9}
\simeq 0.570044\,\Lambda^3,
\eea
We see that one of the color directions is preferred. In this case it is the 33
direction as we have aligned $A_4$ at spatial infinity along the third axis. 
Contrary to $<A_4^3>$ which vanishes in the decompactified limit as $1/l$
the difference in the color components of the gluino condensate remains finite 
(and computable) in the strong-coupling limit. \\

\Eqs{12}{3} demonstrate the dynamical Higgs effect (as there are no elementary 
Higgs fields in the pure SYM theory). In this case, the colored gluino condensate
$<\!\Lambda^{ab}\!>$ (the composite Higgs field) belonging to the dimension-5 traceless
symmetric tensor representation has a nonzero v.e.v. \ur{chi} that breakes $SU(2)$ 
down to the $U(1)$ subgroup in the same sense as the v.e.v. of an elementary 
Higgs field does. \\


\section{Discussion}

The appearance of a nonzero color gluino condensate is, in a sense,
trivial. In partially compactified $R^3\times S^1$ SYM theory the gauge
group is apparently spontaneously broken to the maximal Abelian
subgroup by dyons, at least when the compact dimension is much less 
than the $\Lambda$ scale of the theory.  The minimum of the superpotential 
induced by dyons corresponds to a nonzero $A_4$ \cite{DHKM} which 
has to lie in some direction in color space thus breaking the color group. 
Therefore, in the compactified pure SYM theory a Higgs effect 
takes place, with $A_4$ playing the role of the Higgs field in the 
adjoint representation. Unpleasantly, $A_4$ is not Lorentz-invariant, 
but the effect is there. The non-zero value of the colored gluino condensate
$<\!\Lambda^{ab}\!>$ is a Lorentz-invariant manifestation of the same 
symmetry breaking. \\

What is interesting, the value of the color gluino condensate we
have found does not depend on the compactification circumference $l$
-- just as the `normal' gluino condensate is independent of $l$;
they are holomorphic quantities. Therefore, one can claim that color 
symmetry remains broken in the decompactified strong-coupling limit. 
An indirect evidence for color symmetry breaking follows
from the fact that the correct value of the color-singlet gluino
condensate is obtained from field configurations with a non-trivial
holonomy. Strong-coupling instantons have a trivial holonomy, 
and they produce only a color-singlet condensate -- in contradiction 
with \eq{LaLa} following merely from supersymmetry. But even the 
color-singlet condensate gets a wrong value from instantons. 
We have shown in section 3 that to get the correct value it is 
insufficient to add other field configurations on top of the instanton: 
one has to replace the instanton by the caloron with non-trivial holonomy, 
in the first place. A non-trivial holonomy means that there is a privileged 
direction in color space, which is equivalent to color symmetry breaking.  \\

An additional although so far indirect demonstration of the
importance of dyons at strong coupling has come very recently from
another end. Using wrapped $D5$ branes \cite{DVLM} or warped
deformed conifold \cite{I} the authors obtained the correct all-loop
$\beta$ function of ${\cal N}=1$ SYM theory from corresponding
supergravity solutions. In addition, both references find the same
non-analytic corrections to the $\beta$ function, which are naturally
associated with the pairs of dyons, not instantons. \\

Dyons have long-range Coulomb interactions which are Debye-screened
in the plasma. It results in magnetic photons getting a mass
and in confinement {\it {\`a} la} Polyakov \cite{Pol}. Polyakov's
scenario of confinement in $3d$ is essentially {\em Abelian}. It
implies that the gauge group is spontaneously broken to the maximal
Abelian subgroup, that the `charged' gluons get masses via the Higgs
mechanism but are confined, and that the Abelian magnetic `photons'
get mass from Debye screening. Qualitatively, the same Abelian
scenario has been discussed for the $4d$ pure gauge theory by
't Hooft \cite{tH}. It can be made quantitative in the weak-coupling
regime of the $4d$ ${\cal N}=2$ theory softly broken to ${\cal N}=1$
\cite{SW,DS}. The spontaneous breaking of the gauge group to 
the maximal Abelian subgroup is a welcome feature: the Abelian 
confinement is well understood, at least on the philosophical level, 
and has a chance to be ultimately put into a quantitative form following the
lines of the references cited above. \\

The Abelian scenario has clear signatures in the weak-coupling regime
but they become not so clear in the strong-coupling limit, especially
if the theory confines color and only gauge-invariant correlators are 
the observables. Strictly speaking, there is no gauge-invariant local
order parameter which would distinguish between Abelian
confinement and a `true' non-Abelian case. To find out 
uniequivocal observable signatures of the color-broken scenario and to 
check whether the pure (not supersymmetric) YM theory has the 
same features is an intriguing task which we postpone for the future.\\ 

\newpage
\noindent {\large\bf Acknowledgements} \\

\noindent
We are grateful to Gennady Danilov, Alexander Gorsky, Alex Kovner, 
Alexei Yung and Ariel Zhitnitsky for useful discussions.  
V.P. acknowledges partial support from the RFBR grant No. 00-15-96610.  
Both of us are grateful to Klaus Goeke and Maxim Polyakov for the hospitality 
at Bochum University where part of this work has been done, and for a 
support from Deutsche Forschungsgemeinschaft and Sofia Kovalevskaya grants.

\def\appendixes{\par\setcounter{section}{0}
\setcounter{subsection}{0}
\setcounter{equation}{0}   
\def\thesection{Appendix.  
}
\def\theequation{\Alph{section}.\arabic{equation}}}
\appendixes

\section{L,M monopoles}

We give here explicit expressions for the fields
of the four dyons in the compactified $R^3\times S^1$ space,
with all four possible signs of the electric and magnetic charges.
The two usual Bogomolny--Prasad--Sommerfeld (BPS) dyons in the regular
(`hedgehog') gauge have the form:

\bea
\la{A41}
A_4^a &=& \mp n_a\,v\Phi(vr),\qquad \Phi(z)=\coth z -\frac{1}{z}\quad
\stackrel{z\to\infty}{\longrightarrow}\quad 1-\frac{1}{z}+O(e^{-z}),\\
\nonumber \\
\la{Ai1}
A_i^a &=& \epsilon_{aij}n_j\,\frac{1-R(vr)}{r},
\qquad R(z)=\frac{z}{\sinh z}\quad \stackrel{z\to\infty}{\longrightarrow}
\quad O(ze^{-z}).
\eea
Here $r=\sqrt{x_1^2+x_2^2+x_3^2}$, $n_a=x_a/r$. The upper sign in $A_4$
corresponds to the self-dual ($E^a_i=F^a_{i4}=B^a_i=\half\epsilon_{ijk}
F^a_{jk}$) and the lower sign to the anti-self-dual ($E^a_i=-B^a_i$)
solution. We shall call them $M$- and $\bar M$-monopoles, respectively.\\

The magnetic field strength in the hedgehog gauge is given
by two structures:

\bea
\la{B0}
B^a_i &=& (\delta_{ai}-n_an_i)\,F_1(r)+n_an_i\,F_2(r),
\qquad{\rm where} \\
\nonumber \\
\la{F1}
F_1(r) &=& \frac{1}{r}\frac{d}{dr}R(vr)=-v\frac{R(vr)\Phi(vr)}{r}
=\frac{v^2}{\sinh(vr)}\left(\frac{1}{vr}-\coth(vr)\right)
=v^2O\left(e^{-vr}\right),\\
\nonumber \\
\la{F2}
F_2(r) &=& -\frac{d}{dr}v\Phi(vr)=\frac{R^2(vr)-1}{r^2}
=\frac{v^2}{\sinh(vr)}-\frac{1}{r^2}
=-\frac{1}{r^2}+v^2O\left(e^{-2vr}\right).
\eea

If there is more than one monopole in the vacuum it is impossible to
add them up in the hedgehog gauge: one has to ``gauge-comb'' them to
a gauge where $A_4^a$ has the same asymptotic value at spatial
infinity for all monopoles involved, -- say, along the third color
axis. It is achieved with the help of two unitary matrices dependent
on the spherical angles $\theta,\phi$:

\bea
\la{Sp}
S_+(\theta,\phi) &=& e^{-i\frac{\phi}{2}\tau^3}
e^{i\frac{\theta}{2}\tau^2}e^{i\frac{\phi}{2}\tau^3},\qquad
S_+(n\cdot \tau)S_+^\dagger=\tau^3,\\
\nonumber \\
\la{Sm}
S_-(\theta,\phi) &=& e^{i\frac{\phi}{2}\tau^3}
e^{i\frac{\pi-\theta}{2}\tau^2}e^{i\frac{\phi}{2}\tau^3},\qquad
S_-=-i\tau^2S_+,\qquad S_-(n\cdot \tau)S_-^\dagger=-\tau^3.
\eea

We shall gauge-transform the $M$-monopole field with $S_-$ and the
$\bar M$-monopole with $S_+$. As the result their $A_4$ components
become equal:

\beq
A_4^{M,\bar M}=v\Phi(vr)\frac{\tau^3}{2}=\left[v-\frac{1}{r}
+O\left(e^{-vr}\right)\right]\frac{\tau^3}{2}.
\la{A4M}\eeq
On the contrary, the spatial components differ in sign. We write
them in spherical components:

\beq
\pm A_i^{M,\bar M}=\left\{\begin{array}{c}
A_r = 0 \\
\nonumber \\
A_\theta = \frac{R(vr)}{2r}(\tau^1\,\sin\phi+\tau^2\,\cos\phi)\\
\nonumber \\
A_\phi = \frac{R(vr)}{2r}(\tau^1\,\cos\phi-\tau^2\,\sin\phi)
+\frac{1}{2r}\tan\frac{\theta}{2}\tau^3.
\end{array}\right.
\la{Ai2}\eeq

The azimuthal component of the gauge field has a singularity along
the negative $z$ axis, therefore we shall call it the stringy gauge.
The field strength, however, has no singularities. The electric field
both of $M$ and $\bar M$ monopoles in the stringy gauge is

\beq
E_i^{M,\bar M}=\left\{\begin{array}{ccc}
E_r = - \frac{F_2(r)}{2}\tau^3
\quad \stackrel{r\to\infty}{\longrightarrow}
\quad \frac{1}{r^2}\frac{\tau^3}{2} \\
\nonumber \\
E_\theta = \frac{F_1(r)}{2}(-\tau^1\,\cos\phi+\tau^2\,\sin\phi)
=v^2 O\left(e^{-vr}\right)\\
\nonumber \\
E_\phi = \frac{F_1(r)}{2}(\tau^1\,\sin\phi+\tau^2\,\cos\phi)
=v^2 O\left(e^{-vr}\right)
\end{array}\right.
\la{EiM}\eeq
while the magnetic field is $B_i=\pm E_i$. Therefore, $M$ monopole
has (electric, magnetic) charges $(++)$ whereas the $\bar M$ one has
$(+-)$. \\

There is a second pair of dyons \cite{LY}: a self-dual one with the
charges $(--)$ which we shall name $L$-monopole, and an
anti-self-dual one with charges $(-+)$ which we shall name $\bar L$
monopole. They are obtained from eqs.\ref{A41}-\ref{F2} by replacing
$v\to \frac{2\pi}{l}-v$.  One first transforms them from the hedgehog
to the stringy gauge with the help of the unitary matrices $S_+$ and
$S_-$, respectively. As the result, they get the same asymptotics
$A_4(\infty)=\left(-\frac{2\pi}{l}+v
+\frac{1}{r}\right)\frac{\tau^3}{2}$. To put the asymptotics in the
same form as for $M,\bar M$-monopoles (see \eq{A4M}) one makes an
additional gauge transformation with the help of the time-dependent
matrix

\beq
U=\exp\left(-i\frac{\pi}{l}x^4\tau^3\right).
\la{U}\eeq
This gives the following fields of $L,\bar L$ monopoles in the
stringy gauge:

\beq
A_4^{L,\bar L}=\left[\left(\frac{2\pi}{l}-v\right)
\Phi\left(\left|\frac{2\pi}{l}-v\right|r\right)-\frac{2\pi}{l}\right]
\frac{\tau^3}{2}\quad \stackrel{r\to\infty}{\longrightarrow}
\quad =\left(v+\frac{1}{r}\right)\frac{\tau^3}{2},
\la{A4L}\eeq

\bea
\la{EiL}
E_i^{L,\bar L} &=& \left\{\begin{array}{ccc}
E_r = \frac{F_2(r)}{2}\tau^3
\quad \stackrel{r\to\infty}{\longrightarrow}
\quad -\frac{1}{r^2}\frac{\tau^3}{2} \\
\nonumber \\
E_\theta = -\frac{F_1(r)}{2}\,U(x^4)
(-\tau^1\,\cos\phi+\tau^2\,\sin\phi)U^\dagger(x^4)\\
\nonumber \\
E_\phi =-\frac{F_1(r)}{2}\,U(x^4)
(\tau^1\,\sin\phi+\tau^2\,\cos\phi)
U^\dagger(x^4),\end{array}\right. \\
\nonumber \\
\la{BiL}
B_i^{L,\bar L} &=& \pm E_i^{L,\bar L}.
\eea

The `profile' functions $F_{1,2}$ are given by \eqs{F1}{F2}, with
the replacement $v\to \frac{2\pi}{l}-v$. We notice that ``the
interior'' of the $L,\bar L$ dyons, represented by the $\theta,\phi$
field components are time dependent. This is why in the true $3d$
case these objects do not exist. \\

The properties of the four dyons are summarized in Table 1. \\

\begin{table}
\begin{center}
\begin{tabular}{|c|c|c|c|c|}
\hline
&&&& \\
                & $M$ & $\bar M$ & $L$ & $\bar L$  \\
&&&& \\
\hline
&&&& \\
 electr. charge      &  +  &  +     &  $-$   &   $-$      \\
&&&& \\
 magn. charge    & + &  $-$       &   $-$  &   +      \\
&&&& \\
 action, $\frac{4\pi l}{g^2}$ & $v$ & $v$ & $\frac{2\pi}{l}-v$ &
$\frac{2\pi}{l}-v$  \\
&&&& \\
top. charge & $+\half$ & $-\half$ & $+\half$ & $-\half$  \\
&&&& \\
\hline
\end{tabular}
\end{center}
\caption{Four dyons of $SU(2)$.}
\end{table}

\end{document}